\documentclass[useAMS,usenatbib]{mn2e}
\usepackage{graphicx}
\usepackage{color}
\newcommand{\msun}{\ensuremath{\rmn{M}_{\sun}}} 
\newcommand{\kms}{\ensuremath{\rmn{km}~\rmn{s}^{-1}}} 

\title[Simulating Sinking Satellites with {\sc Superbox-10}]
{Simulating Sinking Satellites with {\sc Superbox-10}}
\author[R. Bien, T. Brandt, A. Just]{R. Bien\thanks{E-mail: reinhold@ari.uni-heidelberg.de}, 
T. Brandt, A. Just\thanks{E-mail: just@ari.uni-heidelberg.de}\\
Astronomisches Rechen-Institut, Zentrum f\"ur Astronomie, 
University of Heidelberg, M\"onchhofstr. 12-14, 69120 Heidelberg, Germany}
\begin{document}

\date{Accepted 2012 October 3. Received 2012 July 27; in original form 2012 January 23}

\pagerange{\pageref{firstpage}--\pageref{lastpage}} \pubyear{2012}

\maketitle

\label{firstpage}

\begin{abstract}
{\sc Superbox-10} is the successor of {\sc Superbox}, a particle-mesh code 
where additional grids and sub-grids are applied to regions of high particle density. 
Previous limitations have been solved. For instance, the vertical resolution is
improved considerably when flattened grids are used.
Since the computationally most intensive part is the Fast Fourier Transform, we
introduce a parallelised version using the library {\sc fftw}, resulting in a 
speed-up of a few.
The new features are tested using a galaxy model consisting of an exponential disc,
a bulge and a dark matter halo. We demonstrate that the use of flattened grids
efficiently reduces numerical heating.
We simulate the merging of disc-bulge-halo galaxies with small spherical satellites.
As a result, satellites on orbits with both low eccentricity and inclination heat
the disc most efficiently. Moreover, we find that most of the satellite's energy
and angular momentum is transfered to the halo.

\end{abstract}


\section{Introduction}

Observations of stars in the solar neighbourhood show an increase of the
velocity dispersion (the random motion) proportional
to $t^{0.3\dots 0.6}$, where $t$ is the stellar age (Wielen 1977; Holmberg, Nordstr\"om \& Andersen 2007). 
This effect can be interpreted as a 
heating of the Galactic disc with time.
However, it is unclear what is the cause of this heating. Several mechanisms have been proposed:
Massive black holes in the galactic halo as a source of heating were 
investigated by \cite{lo85}, \cite{wf90}, \cite{rl93} and \cite{hf02}, 
but seem to be
excluded by observations. \cite{atb03} considered
massive clumps of dark matter as a possible heating agent. Giant molecular
clouds were found to cause mostly vertical heating by \cite{l84} and later
\cite{hf02}. \cite{cs85} and \cite{c87} show that transient spiral waves
heat efficiently in the Galactic plane. Lastly, \cite{qhf93} and \cite{vw99} simulate
merging with small satellites and find both radial and vertical heating. This
last mechanism is discussed in this paper. It should be mentioned, however, that
radial mixing may influence the age-velocity dispersion relation significantly
(Sch\"onrich \& Binney 2009; Sellwood \& Binney 2001; Ro\v{s}kar et al. 2008).

The merging of galaxies can be divided into three categories according to their mass ratios.
Major mergers with mass ratios 1:1 to 1:3 of the total galaxy masses usually destroy discs
and result in an early type remnant \citep[][and references therein]{njb06}. Minor mergers
with mass ratios 1:3 to 1:10 of the total galaxy masses destroy a thin disc \citep{pbk10}. 
The minor merger events are discussed in the context of the formation of thick discs
\citep{vh08,vh09} and substructure in the halo \citep{n02,pbk10}. The third category is the merging
of satellite galaxies (or dark matter clumps) with
smaller mass ratios. These merger or interaction events can be characterized as a
perturbation of the primary galaxy and it is more convenient to quantify the mass ratio in units
of the disc mass of the primary. In this context the survival of a thin disc over 10 Gyr in a
$\Lambda\rmn{CDM}$ universe lead to serious constraints on the clumpiness of the DM halo \citep{atb03}. 
\cite{h08} compared the efficiency of the disc by
satellite galaxy mergers of different authors and argued that it scales with the square of the
satellite mass. This is consistent with the interpretation heating by dynamical friction during
each disc crossing event. As a consequence, the upper mass end of the perturbers are most
efficient and determine the survival of a thin disc. Kazantzidis et al. (2008, 2009) performed a
comprehensive study of the impact of DM subhaloes on a Milky Way like disc galaxy in a
realistic cosmological context. They investigated the resulting substructures and kinematic
features in disc and halo.

In numerical simulations the long-term (or secular) dynamical evolution of a thin galactic disc
is very sensitive to the spatial resolution, the level of numerical noise, and the dynamical
feedback with the DM halo. A live dark matter halo is important for two reasons. \cite{vw99} 
have shown that the reaction of the disc on the perturbation of a satellite
galaxy is significantly higher with a live halo compared to a rigid dark matter potential.
Secondly, dynamical friction in the dark matter is important for the orbital evolution and thus
energy loss of satellites with masses exceeding $\sim 10^8~\msun$. This is exactly the mass range
dominating the heating rate of the disc by satellite galaxies. Additionally, the mass of the dark
matter particle must not be too large in order to avoid numerical heating of the disc. In
most simulations the disc is represented by less than 1 million particles leading to a significant
thickening of the disc by numerical two-body relaxation, which is much stronger in direct $N$-body 
codes and Tree codes compared to particle mesh-codes. In \cite{vw99} and in \cite{hc06} 
the numerical heating of the unperturbed disc was
shown explicitly. \cite{vw99} used a differential method to quantify the
dynamical heating of infalling satellites. This may underestimate the heating rate, since a
more realistic thinner, dynamically cooler, disc is more sensitive to perturbations. On the other
hand, the feedback on the satellite galaxy may be larger by the stronger disc shocking event.
Particle-mesh codes are predestined for the simulation of collisionless systems such as
galaxies, because two-body relaxation is strongly suppressed by the partial decoupling of the
point-like structure in orbital motion of the particles and the gravitational potential based on
the grid. \cite{k07} have shown that the growth of global mode perturbations
are easily damped out by noise due to an insufficient particle number. If the excitation of spiral
structure or bar-like perturbations are an important mechanism for the energy transfer from
the satellite to the disc, then a very high resolution well above 1 million particles is needed to
reproduce a realistic heating rate.

Improving the resolution of discs in vertical direction has always been
a challenge.
There is, indeed, a further problem which is inherent in any numerical
technique that is used for the simulation of disc galaxies. Stellar discs
have relaxation times larger than a Hubble time, i.\,e. they are collisionless
systems. This implies that in simulations the thickness of an unperturbed 
disc indicates how well the code would model a collisionless gravitating 
system.
The term ``unperturbed'' denotes the absence of perturbations such as 
stellar bars, spiral arms, or molecular clouds.
One often notes, in simulations, that discs become thicker with time. 
This is caused
by the graininess of the particle distribution and the limited spatial
resolution. We refer to this effect as ``numerical heating''.

For two decades {\sc Superbox} is used as a successful tool to simulate 
the dynamics of isolated and interacting galaxies. 
A first description was given by Bien, Fuchs \& Wielen (1991), 
together with a consideration of the direct $N$-body technique and a 
tree-code. For a more detailed discussion the reader is referred to \cite{f00}. 
We also mention Fellhauer's lecture notes \citep{fln08}.
The code was evolving from the conventional particle-mesh technique 
\citep[see, e.\,g.,][]{he88}
by applying grids and sub-grids in regions of high particle 
density. The advantage of {\sc Superbox} is that the code can run on any 
workstation or PC, giving reliable results. The code thus proves to be a 
serious alternative to direct $N$-body schemes and tree-codes.

Initially, \cite{mb93} applied {\sc Superbox} to compare 
numerical experiments to observations of the high-velocity encounter of 
NGC 4782/4783. They found a convincing description of the morphological and 
kinematical structure, and estimated the time elapsed since the closest 
approach of the two interacting galaxies. This was the beginning 
of a series of research projects using {\sc Superbox} as a tool. We note
as examples the dynamical evolution of a satellite ga\-la\-xy 
\citep{kk98}, the decay of dwarf ga\-la\-xies in dark 
matter haloes \citep{p02}, and dynamical friction, compare
\cite{p04} and \cite{jp05}.  It should be mentioned that
\cite{s03} studied the inspiral of a black hole to the Galactic 
centre, and \cite{k07} considered unstable modes in a
disc. \cite{f08} published on the dynamics of the 
Bootes dwarf galaxy. \cite{p08} investigated the tidal 
evolution of dwarfs of the Local Group. Most recently, a study was
made on the dynamical friction of massive objects in galactic centres by \cite{j11}. 
Our list is by far not complete.

So far, {\sc Superbox} was not able to adapt the sub-grids to discs.
A higher resolution in vertical direction can be
achieved by introducing flattened grids, see \cite{b08}. This 
new feature and the application to dynamical heating of galactic
discs are the main topics of the present paper. The improved code 
is called {\sc Superbox-10}.
As {\sc Superbox}, {\sc Superbox-10} is rather free of numerical heating.

\begin{figure}
\vspace{0.0cm}
~
\hspace{-0.80cm}
\includegraphics[width=80.00mm,height=73.06mm]{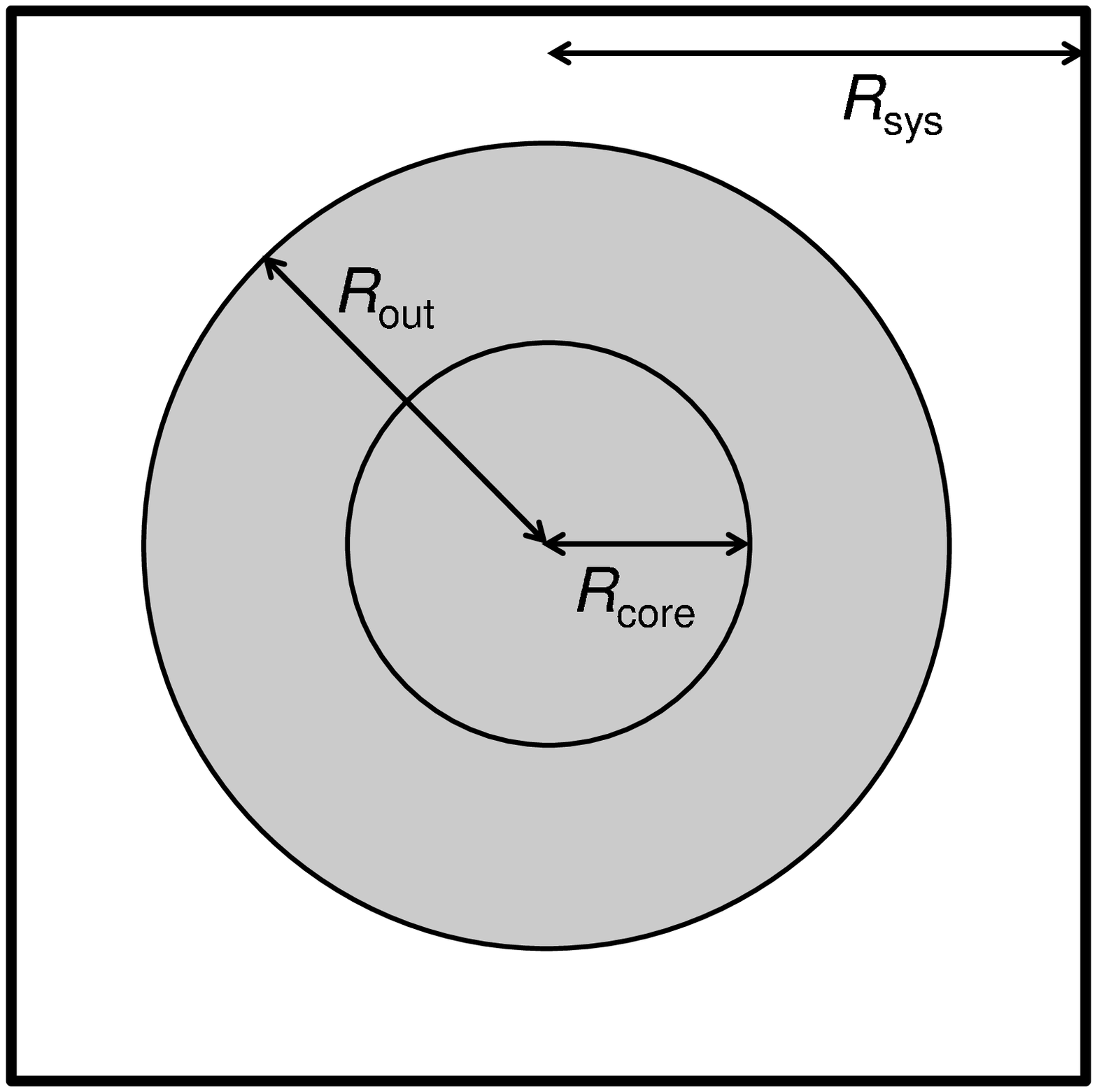}
~
\vspace{0.25cm}
\caption{Projection of a simplified galaxy model in a 3D box. Shown are the basic parameters $R_{\rmn{core}}$,
$R_{\rmn{out}}$ and $R_{\rmn{sys}}$.}
\label{label1}
\end{figure} 

Below we first describe {\sc Superbox} in detail and then extend to {\sc Superbox-10}.
It follows a discussion of the parallelised version which results in a significant speed-up.
Then our test of improved vertical resolution is described. As an important 
application we consider disc heating by satellite galaxies. In particular, we address
the ratio of radial and vertical velocity dispersion and the transfer of energy and
angular momentum. Special attention is given to the interaction of the infalling
satellite with the dark matter halo.

  In a conference proceedings publication \citep{b08} we presented preliminary results
  on an improved version of {\sc Superbox} with increased vertical resolution of the disc:
Negligible disc thickening of an isolated disc after 2.5 Gyr. The present paper should be seen
as a pilot study for high resolution simulations of disc heating by satellite galaxy mergers.
Two new aspects are presented: (1) No correction for numerical heating is needed and we
can measure the heating rate directly. Dynamical friction in the dark matter halo is included
automatically. (2) We analyse for the first time additionally the transfer of angular momentum.
This is important for understanding the physical processes responsible for the disc heating.
For this pilot study we select a satellite with a relatively high mass on eccentric orbits, which is
destroyed after a few disc crossing events.

\section[]{A review of {\sc Superbox}}

\subsection{Basic concepts}

The conventional particle-mesh technique considers a set of massive particles,
often called ``superstars'', in a three-dimensional Cartesian grid (``box'') consisting 
of $N \times N \times N$ cells which represents a relevant part of the universe.
In this section, the length of each cell is the unity.
We suppose that $\rho_{a,b,c}$ is the mean density in the cell with indices
$a$, $b$, $c$ which are integers running from 0 to $N-1$. Then the potential
can be obtained by solving Poisson's equation
\begin{equation}\label{eq:poisson}
  \Phi_{a,b,c} = G \sum_{\xi = 0}^{N-1} \sum_{\eta = 0}^{N-1} \sum_{\zeta = 0}^{N-1}
\rho_{\xi,\eta,\zeta} H_{\xi-a,\eta-b,\zeta-c}
\end{equation}
Here, $H_{\xi-a,\eta-b,\zeta-c}$ (i.\,e. Green's function) is the inverse distance
between the points having indices $a$, $b$, $c$ and $\xi$, $\eta$, $\zeta$, namely
\begin{equation}
\label{eqn:greens}
  H_{\xi-a,\eta-b,\zeta-c} = [(\xi-a)^2 + (\eta-b)^2 + (\zeta-c)^2]^{-1/2}
\end{equation}
We note that the term $G \rho_{\xi,\eta,\zeta} H_{\xi-a,\eta-b,\zeta-c}$ 
accounts for the contribution of the cell $\xi$, $\eta$, $\zeta$ to the
cell $a$, $b$, $c$. Thus $\Phi_{a,b,c}$ gives the correct potential 
assigned to cell $a$, $b$, $c$. The evaluation of all $N \times N \times N$ values 
$\Phi_{a,b,c}$ takes a time proportional to the number of cells squared, 
$(N \times N \times N)^2$.

\begin{figure}
\vspace{0.0cm}
~
\hspace{-1.0cm}
\includegraphics[width=90.00mm,height=82.19mm]{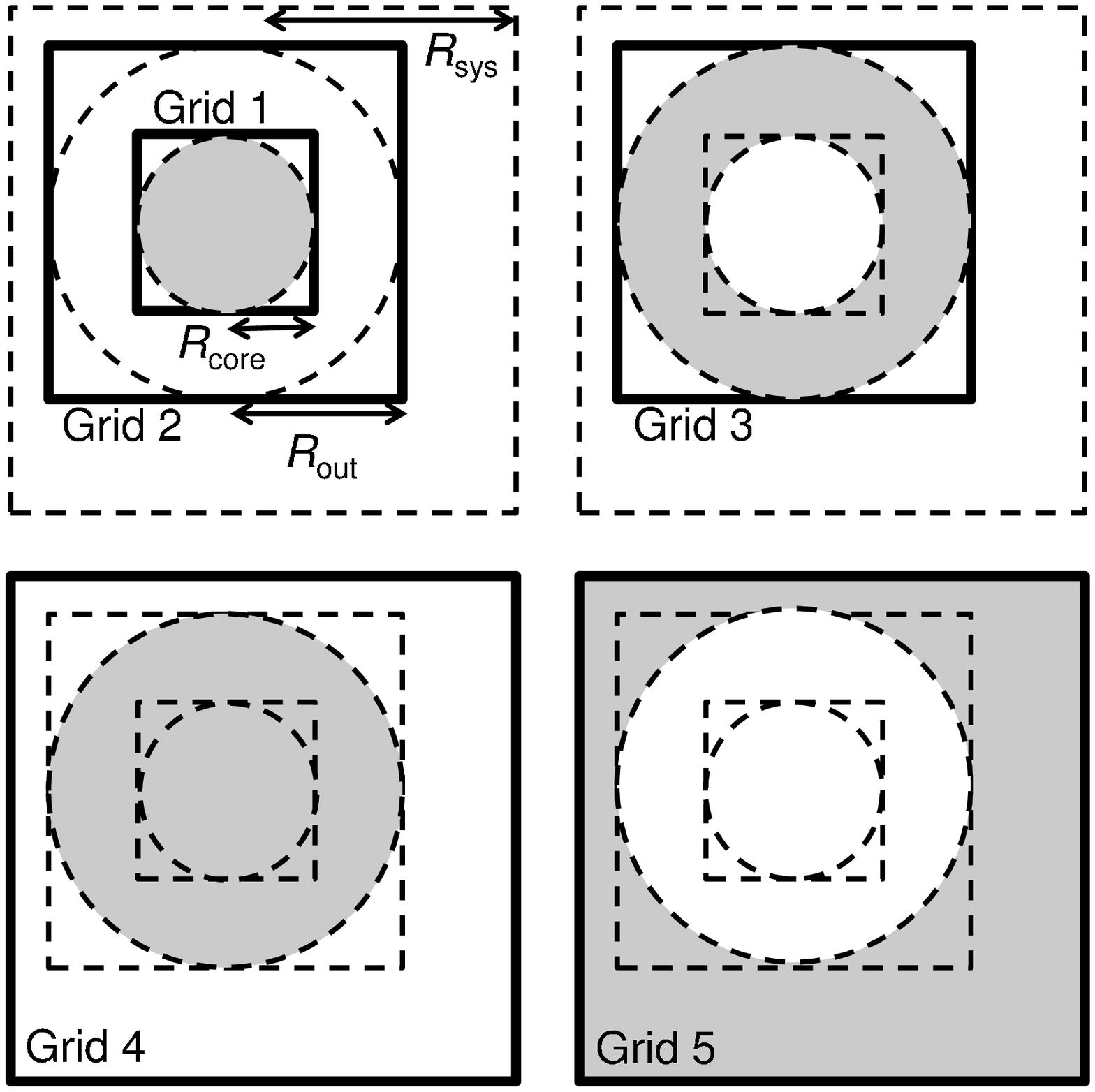}
~
\vspace{0.25cm}
\caption{The five grids of {\sc Superbox}. Grid 1 has the highest resolution
and resolves the core. Grid 1, grid 2 and grid 3 move through the local universe,
defined by grid 4 and grid 5.}
\label{label2}
\end{figure} 

The direct convolution can be replaced by discrete Fourier transforms where $N$
is supposed to be a power of 2. First $\rho_{a,b,c}$ and $H_{a,b,c}$ are
transformed, resulting in
\begin{equation}
  \hat{\rho}_{k,l,m} = \sum_{a = 0}^{N-1} \sum_{b = 0}^{N-1} \sum_{c = 0}^{N-1}
  \rho_{a,b,c} \exp[-i{2\pi \over N}(ka+lb+mc)]
\end{equation}
and $\hat{H}_{k,l,m}$, analogously. The potential is then obtained by the
inverse transformation of the product $\hat{\rho}_{k,l,m} \hat{H}_{k,l,m}$,
\begin{eqnarray}
\Phi_{a,b,c} = \nonumber\\
{G \over N^3} \sum_{k = 0}^{N-1} \sum_{l = 0}^{N-1} \sum_{m = 0}^{N-1}
\hat{\rho}_{k,l,m} \hat{H}_{k,l,m} \exp[+i{2\pi \over N}(ka+lb+mc)]  
\end{eqnarray}
The procedure can dramatically be accelerated when a Fast Fourier Transform
(FFT) is applied. The computing time is then roughly proportional to
$N^3\log N$.

The potential, as derived so far by Fourier transforms, is correct for
periodic systems only. The exact potential of isolated systems can be obtained
by doubling the number of cells in each coordinate direction, i.\,e. a
grid is considered which contains $2N \times 2N \times 2N$ cells. 
A rigorous proof can be found in the paper by \cite{eb79}.
As before,
the $N \times N \times N$ sub-grid contains the particles (``active region'').
The remaining space is left empty. Likewise, $H$ is extended over the
$2N \times 2N \times 2N$ cells such that it satisfies
\begin{eqnarray}\label{eq:symmetry}
H_{2N-a,b,c} = H_{2N-a,2N-b,c} \nonumber\\ 
= H_{2N-a,b,2N-c} = H_{2N-a,2N-b,2N-c} = H_{a,2N-b,c} \nonumber\\
= H_{a,2N-b,2N-c} = H_{a,b,2N-c} = H_{a,b,c} 
\end{eqnarray}
for $0 \leq a,b,c \leq N$. Having transformed the extended functions 
$\rho$ and $H$, the correct potential $\Phi$ can be found in the active
grid. Outside, $\Phi$ is unphysical. 

Numerical differentiation of the potential with respect to the coordinates
leads for each particle $j$ to the acceleration $(a_{x})_j$, $(a_{y})_j$,
 $(a_{z})_j$. The leap-frog scheme is applied to integrate the
equations of motions. In $x$-direction the algorithm reads
\begin{eqnarray}
\dot{x}_j(t+ \Delta t/2) &=& \dot{x}_j(t- \Delta t/2) + \Delta t (a_{x})_j \nonumber \\
x_j(t+ \Delta t) &=& x_j(t) + \Delta t \dot{x}_j(t+ \Delta t/2) 
\end{eqnarray} 
where $t$ is the time and $\Delta t$ denotes the constant integration step.

The particle-mesh technique holds some subtle twists. 
For instance for $H_{a,b,c}$ it suffices a $(N+1) \times (N+1) \times (N+1)$
grid. From the symmetry conditions (\ref{eq:symmetry}) follows that the function is then
known on the whole $2N \times 2N \times 2N$ grid. The sub-grid is overwritten
by $\hat{H}$ which shows similar symmetries. $\hat{H}$ is calculated only once,
since it is fixed over the integration period. For each integration step the grid
containing the density $\rho$ is overwritten by the Fourier transform $(\hat{\rho}$ 
which, in turn, is overwritten by $\hat{\rho}\hat{H}$). This product is finally
overwritten by $\Phi$. The advantages of the method are obvious. 
The conventional particle-mesh technique was 
orders of magnitude faster than direct $N$-body codes at the time. This allows the
integration of a large number of particles. Thus, statistical noise is extremely
low. Particles leaving the outermost grid are lost. The user should thus carefully consider the number of
particles involved in the computation. 

The greatest disadvantage, however, is the low spatial resolution in regions
of higher particle densities, e.\,g. the cores of spherical galaxies. An
improvement is {\sc Superbox} which treats the particles self-consistently
in a nested system of grids. Here, the term ``self-consistent'' means that
all particles have masses which define the total potential. For simplicity,
we consider a spherical galaxy of radius $R_{\rmn{out}}$. Fig. 1 shows how the 
particles are subdivided into a central region (``core'') where the distance 
$r$ from the centre is $< R_{\rmn{core}}$ and a region with 
$R_{\rmn{core}} < r < R_{\rmn{out}}$. The local universe.is
given by $R_{\rmn{sys}}$. 

As shown in Fig. 2 {\sc Superbox} utilises five grids. They all have the same number
of cells, $N^3$. 
\begin{enumerate}
  \item Grid 1 has the highest resolution and resolves the core which contains
        all particles with $< R_{\rmn{core}}$.         
  \item Grid 2 contains the core, too, but the resolution is intermediate.
  \item Grid 3 is equal to grid 2, but contains only particles with
        $R_{\rmn{core}} < r < R_{\rmn{out}}$.  
  \item Grid 4 is the fixed global grid. It contains only particles inside
        $R_{\rmn{out}}$.  
  \item Grid 5 is equal to grid 4 with all particles outside $R_{\rmn{out}}$. 
\end{enumerate}
As the galaxy moves, grid 1, grid 2 and grid 3 move through the local universe
(i.\,e. grid 4 and grid 5, respectively). This is done by centreing the inner and intermediate
grids on the density maximum of the galaxy in question. Alternatively, the centre of mass
can be considered. Particles outside $R_{\rmn{sys}}$
are lost, compare the discussion above. The code is not restricted to one galaxy. Since the potentials, and
thus the accelerations, are additive, all galaxies are treated sequentially
in the same five grids. The corresponding five total potentials are $\Phi_1$, 
$\Phi_2$, $\Phi_3$, $\Phi_4$, and $\Phi_5$  
\begin{enumerate}
  \item For a particle with $r < R_{\rmn{core}}$, the correct potential is
        $\Phi_1 + \Phi_3 + \Phi_5$.  
  \item For a particle with, $R_{\rmn{core}} < r < R_{\rmn{out}}$ the sum
        $\Phi_2 + \Phi_3 + \Phi_5$ is taken.
  \item When the particle is outside $R_{\rmn{out}}$ then $\Phi_4$
        and $\Phi_5$ are used.    
\end{enumerate}
In principle, number and type of the galaxies is arbitrary.  

Two features should be emphasized. A particle at the point $(x,y,z)$
is assigned to the cell with indices
\begin{equation}
\label{eqn:cell}
  a=\left[\gamma ~x + {N \over 2}\right], 
    ~ b=\left[\gamma ~y + {N \over 2}\right], 
    ~ c=\left[\gamma ~z + {N \over 2}\right]
\end{equation}
where $[~]$ denotes the nearest integer function. The factor $\gamma$ enhances (or shrinks)
an area. If $\gamma = 10$ is assigned to grid 1, the resolution increases by a factor of 10.
Inside the cell with indices $i, ~ j, ~ k$, let the particle's coordinates be $\Delta x$, 
$\Delta y$, and $\Delta z$. Then its acceleration, e.\,g. in $x$-direction is found by the expression
\begin{eqnarray}
\label{eqn:diff}
a_x = \nonumber\\
{\Phi_{i+1,j,k} - \Phi_{i-1,j,k} \over 2 l_x} \nonumber\\
+{\Phi_{i+1,j,k} + \Phi_{i-1,j,k}  - 2 \Phi_{i,j,k}  \over l_x^2} \Delta x \nonumber\\
+{\Phi_{i+1,j+1,k} - \Phi_{i-1,j+1,k}  +  \Phi_{i-1,j-1,k} - \Phi_{i+1,j-1,k} \over 4 l_x l_y} \Delta y \nonumber\\
+{\Phi_{i+1,j,k+1} - \Phi_{i-1,j,k+1}  +  \Phi_{i-1,j,k-1} - \Phi_{i+1,j,k-1} \over 4 l_x l_z} \Delta z
\end{eqnarray} 
where $l_x = l_y = l_z$ is the length of the cell.
This scheme requires two empty cells at each boundary, i. e.,
only $N-4$ cells per dimension are considered.
We note that in {\sc Superbox}, the
three lengths are equal. When all velocities are updated by applying these accelerations, then new positions
of the particles are calculated. After that, a new integration cycle starts.   

\subsection[]{Extensions to {\sc Superbox-10}}

For version 10, {\sc Superbox}'s original {\sc Fortran 77} code has been ported to 
{\sc Fortran 95} and arranged into several modules, to make future extension easier.

\subsubsection{Individual masses}

The conventional particle-mesh technique considers only a single mass for all particles.
In {\sc Superbox-10}, an individual mass can be assigned to each particle. In practice, 
however, the particles of each subgroup (i.\,e. halo, disc, etc.) carry identical mass.

\subsubsection{Improved vertical resolution}

{\sc Superbox} poorly resolves stellar discs in vertical direction. {\sc Superbox-10} 
overcomes this shortcoming. We explain the basic idea by taking the example of a 
disc-bulge-halo galaxy. First,  
the potential of the disc-bulge-halo galaxy is calculated. The intermediate grids 
grid 2 and grid 3 are flattened along the corresponding $z§$-axis. When, for instance,
the flattening is $q = 1/4$ the resolution is improved by a factor of 4. 
As a consequence, equations \ref{eqn:greens}, \ref{eqn:cell} and \ref{eqn:diff}
need to be changed.
Yet, there is
a restriction to $q$. In order to cover grid 1 and at least two cells of grid 4 (and 5, 
respectively), $q$ should not be smaller than
\begin{equation}
  q_{\rmn{crit}} = \max\left({R_\rmn{core} \over R_\rmn{out}},{4 \over 
{N-4}}{R_\rmn{sys} \over R_\rmn{out}} \right)
\end{equation}
Otherwise, spurious results are obtained. If more than one galaxy is considered, $q < 1$
applies only to the first galaxy.

\cite{b08} simulated a disc-bulge halo galaxy using {\sc Superbox-10}. 
The authors made a comparison with a code based on the TREE-GRAPE scheme \citep{fu05}. 
For 2{,}577{,}235 particles the CPU time turned out to be about three days on a single 
GRAPE6a board. The CPU time of {\sc Superbox-10} (which is still called {\sc Superbox} 
in that paper) on a customary PC is of the same order. This result shows impressively 
the efficiency of the code. \cite{b08} came to the conclusion that the extended code 
\begin{enumerate}
  \item is very fast,
  \item does not produce noticeable numerical disc heating, and
  \item allows an improved vertical resolution.
\end{enumerate}

\subsubsection{Parallelisation}

The computationally most intensive part of {\sc Superbox-10} is the
calculation of the potential using the Fast Fourier Transform.
The non-parallelised version of {\sc Superbox-10} applies the 1D-FFT routine \mbox{\sc realft},
taken from {\it Numerical Recipes} \citep{p92}. The 3D-FFT is calculated
by doing $N\times N$ 1D transforms in each direction.
In the parallelised version, this scheme has been replaced by a 3D-FFT 
from the {\sc fftw} library (version 2.1.5), see \cite{fj05}.
This routine divides the 3D array into slices 
along one dimension and distributes these slices among multiple processors. 
The processors then jointly calculate the 3D-FFT, using the Message Passing
Interface to communicate.
The modular design of {\sc Superbox-10} allows us to replace
the FFT routine without changing much of the rest of the code. The non-parallelised
FFT is still available.

In the following we introduce the fraction $f = t_{\rmn{FFT}}/t_{\rmn{tot}}$ in order
to analyse the non-parallelised version in more detail.
Here $t_{\rmn{FFT}}$ is the time spend in the FFT routine per integration step
and $t_{\rmn{tot}}$ is the total time used for the integration step.
According to
Amdahl's law \citep{a67}, the maximum achievable speed-up by improving the FFT is then given by
\begin{equation}
  S_{\rmn{max}} = \frac{1}{1 - f}
\end{equation}
Table~\ref{tbl:timediv_old} lists $f$, $S_{\rmn{max}}$ and the
speed-up $S$ achieved in benchmarks.

The values depend on the grid size $N$ as well as on the particle 
number $n$. For $n \ga N^3$, the integration of the 
particles' orbits starts to dominate over the potential calculation.
Thus, the higher $N$, the greater the expected speed-up will be.

\begin{table}
\centering
\caption{Maximum and achieved speed-up, $S_{\rmn{max}}$ and $S$, as function of grid
size $N$ and particle number $n$. $f =t_{\rmn{FFT}}/t_{\rmn{tot}}$, see text.}
\label{tbl:timediv_old}
\begin{tabular}{@{}cccccc}
\hline
$N$ & $n$ & $f$ & $S_{\rmn{max}}$ & $S$ & $S/S_{\rmn{max}}$ \\\hline
64 & 500k & 0.62 & 2.61 & 2.48 & 0.95\\\hline
128 & 500k & 0.92 & 12.58 & 8.96 & 0.71\\
& 2M & 0.74 & 3.80 & 3.26 & 0.86\\\hline
256 & 500k & 0.99 & 71.56 & 23.49 & 0.33\\
& 2M & 0.95 & 20.20 & 11.90 & 0.59\\
& 10M & 0.78 & 4.48 & 4.05 & 0.90\\\hline
\end{tabular}
\end{table}

Benchmarks were run on the Titan cluster at the Astronomisches Rechen-Institut, 
which has 32 nodes, each containing a Xeon 3.2 Ghz dual core CPU and a nVidia GeForce 9800 GTX 
Graphics Processing Unit (GPU). 
However, we used only one core per CPU in order to
have more memory available, allowing for higher grid sizes. The GPUs
are currently not used by {\sc Superbox-10}, but a version suitable
for GPUs is in preparation.

\begin{figure}
\centering
\includegraphics[width=0.47\textwidth]{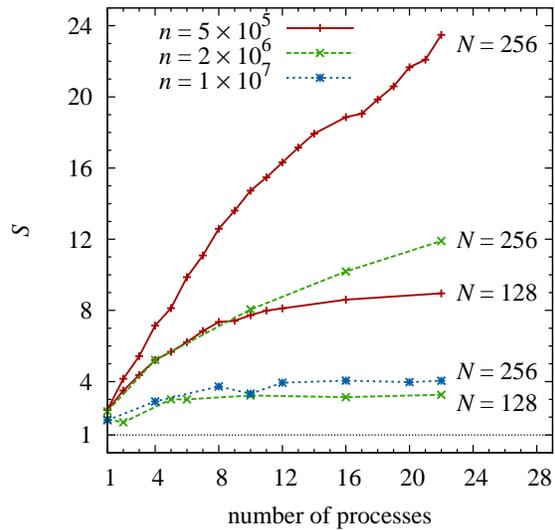}
\caption{Speed-up $S$ compared to non-parallelised {\sc Superbox} as function
of the number of processors for varying $N^3$, the number of grid cells, and particle number $n$.}
\label{fig:speed-up}
\end{figure}

Figure~\ref{fig:speed-up} shows the measured speed-up $S$ as function of the
number of processes in comparison to the
non-parallelised version of {\sc Superbox} with the old FFT when
varying both the grid cell number $N$ and the particle number $n$.
The application of {\sc FFTW} alone already results in
a speed-up between 1.8 and 2.4. In cases where the curves reach their
maximum ($N = 256$ for $n = 10\times10^6$ and $N = 128$ for all $n$)
the speed-up lies between $0.7$ and $0.9~S_{\rmn{max}}$.
In the remaining two cases ($N = 256$ for $n = 0.5\times10^6$ and $n = 2\times10^6$)
it lies between $0.3$ and $0.6~S_{\max}$. When more processors are applied
one can expect a further increase.

The fact that the speed-up almost reaches $S_{\rmn{max}}$ shows that
there is only a small overhead due to communication between the
nodes. Thus, from the values of $S_{\rmn{max}}$ in Table~\ref{tbl:timediv_old} we can formulate the following
rule of thumb: When using $p$ processors and
a grid of size $N^3$, the number of particles should not be greater
than $4 N^3 / p$. Conversely, when simulating $n$ particles with a grid
of size $N^3$, no more than $4 N^3/n$ processors should be used.


\section{Models}

In this section we will describe the models used in our simulations of isolated
disc-bulge-halo galaxies (section \ref{sec:isosim}) and merging of galaxies with satellites
(section \ref{sec:mergesim}).

\subsection{Galaxy model}

Our galaxy model consists of a disc, a bulge and a dark matter halo component. 
The density profile of the disc is exponential in 
radial and isothermal in vertical direction for $R<R_{\max}$ and $|z|<z_{\max}$:
\begin{equation}
  \varrho(R,z) = \varrho_0~e^{-R/h}~\rmn{sech}^2(z/z_0)
\end{equation}

The disc has a scale length $h = 2.5~\rmn{kpc}$ and a thickness $z_0 = 0.6~\rmn{kpc}$.
The profile is cut off at $R_{\rmn{max}} = 10~h$ and $z_{\rmn{max}} = 10~z_0$. 
This excludes $0.05$ per cent of the mass an infinite profile would have. 
At $R = 8~\rmn{kpc}$ the Toomre parameter has a value of $Q = 2$.

Both bulge and halo have a cropped Hernquist profile \citep{h90}
\begin{equation}
  \varrho(r) = \varrho_0~\frac{a^4}{r(r+a)^3}
\end{equation}
with scale radius $a$ and cutoff radius $r_c$.

The scale radius of the bulge is $a = 0.5~\rmn{kpc}$ and is cut off at $r_c = 14~a$.
The halo's scale radius is $a = 16.8~\rmn{kpc}$ and it is cut off at $r_c = 5~a$.

We emphasize that both bulge and halo are represented by live particles. 
The importance of treating the dark matter halo of galaxies as a live component, i.e., 
dynamically evolving and in mutual interaction with the baryonic component, has been
stressed in self-consistent $N$-body simulations of isolated disc galaxies 
(e.g., Athanassoula 2002; Dubinski, Berentzen \& Shlosman 2009). 
The angular momentum exchange between
the disc and dark matter halo in such cases is mediated by dynamical resonances. 
These processes cannot be resolved when using a static, e.g., analytic
prescription for the halo potential.

The model has been implemented with the programme
{\sc MaGaLie} \citep{b01}. We extended {\sc MaGaLie} to allow for up to
10 million particles. 

\begin{table}
\centering
\caption{The three components of our galaxy model and the satellite. 
Shown are the number of particles $n$, the
total mass $M$ and the mass per particle $m$.}
\label{tbl:gal_param}
\begin{tabular}{@{}c|ccc}
\hline
& $n~[10^6]$ & $M~[10^{10}~\msun]$ & $m~[10^3~\msun]$ \\
\hline
disc & 5.19 & 4.82 & 9.30 \\
bulge & 2.30 & 2.14 & 9.30 \\
halo & 2.15 & 20.0 & 93.0 \\
\hline
satellite & 0.50 & 0.54 & 10.08 \\ 
\hline
\end{tabular}
\end{table}

The particle numbers and masses of the three components are listed in Table~\ref{tbl:gal_param}.
In order to reduce the total number of particles,
the particles of the halo are chosen to be ten times as massive as those of disc and bulge
This does not pose a problem though, because they are still light enough not to
cause numerical heating.

Initially, the model is not completely in equilibrium. Therefore,
it is evolved in isolation with high resolution ($N = 256$, $q = 0.25$)
until it reaches equilibrium. This resolution is high enough not to introduce 
any numerical heating.

\begin{figure}
\centering
\includegraphics[width=0.47\textwidth]{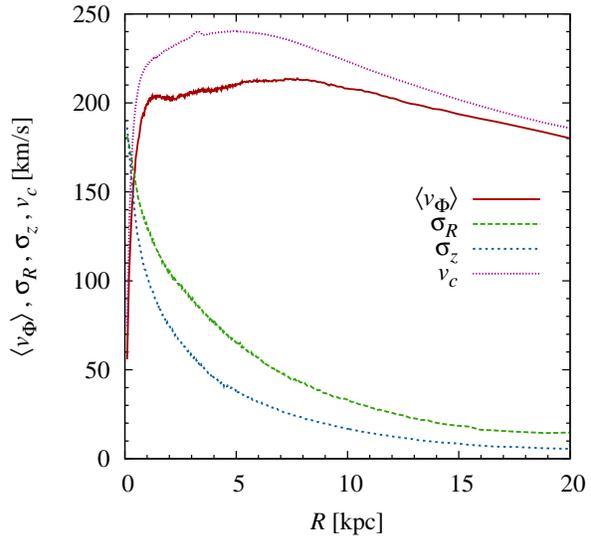}
\caption{The disc's mean rotational velocity $\langle v_{\Phi}\rangle$, 
radial velocity dispersion $\sigma_R$, and vertical velocity dispersion $\sigma_z$.
Also shown is the circular velocity $v_c$. Note that the galaxy has already
reached equilibrium.}
\label{fig:rot-curve}
\end{figure}

Figure~\ref{fig:rot-curve} shows the disc's mean rotational velocity 
$\langle v_{\Phi}\rangle$ and its velocity dispersions $\sigma_R$ and $\sigma_z$ in 
radial and vertical direction after this initial relaxation. Also shown is
the circular velocity $v_c$. The stellar disc has a flat
rotation curve and an exponentially decreasing velocity dispersion.

\subsection{Satellite model}

The satellite for the merger simulations has a Plummer
profile 
\begin{equation}
  \rho(r) = \frac{3M}{4\pi a^3}\left(1 + \frac{r^2}{a^2}\right)^{-\frac{5}{2}}
\end{equation}
with mass $M = 5.4\times10^9~\msun = 0.11\times M_{\rmn{disc}}$ and scale radius $a = 1.5~\rmn{kpc}$. 
Its profile is cut off at $r_c = 10~a$. It consists of $5\times10^5$ particles (see
Table~\ref{tbl:gal_param}).
Like the galaxy, it is first evolved in isolation.


\section{Isolated galaxy models}
\label{sec:isosim}

\cite{b08} showed that {\sc Superbox} intrinsically has a very low level of 
numerical heating. We discuss now in detail the dependence of the numerical
heating on the vertical flattening factor $q$. To that end we simulate the
previously discussed disc-bulge-halo galaxy model in isolation with various 
resolutions and measure the change in vertical thickness of the disc component.

\subsection{Simulation parameters}

\begin{table}
\centering
\caption{Vertical resolution in parsec for various combinations of $q$ and $N$. 
Note that $R_{\rmn{out}} = 28~\rmn{kpc}$.
The radial resolution is independent
of $q$ and has the same value as the vertical resolution for $q=1$.}
\label{tbl:vert_res}
\begin{tabular}{@{}c|cccc}
\hline
$q\backslash N$ & 64 & 128 & 256 & 512 \\
\hline
1 & 933 & 452 & 222 & 110\\
1/2 & 467 & 226 & 111 & 55\\
1/4 & 233 & 113 & 56 & 28\\
1/8 & - & 56 & 27 & 14\\
\hline
\end{tabular}
\end{table}

In all simulations, the grids have 
$R_{\rmn{core}} = 3.5~\rmn{kpc}$, $R_{\rmn{out}} = 28~\rmn{kpc}$ 
and $R_{\rmn{sys}} = 105~\rmn{kpc}$.
Depending on the number of cells $N^3$ and flattening parameter $q$, the
simulations have vertical resolutions listed in Table~\ref{tbl:vert_res}. The radial
resolution for a certain $N=N_0$ has the same value as the vertical resolution
for $N=N_0$ and $q=1$.

We run low-resolution simulations with $N=64$ and $q=1,~0.5,~0.25$. The radial
resolution is 933 pc and the vertical resolution amounts to 933, 467 and 233 pc, respectively.
These simulations are compared to a medium-resolution simulation ($N=128$, $q=0.25$)
where the radial resolution is 452 pc and the vertical resolution is 113 pc.
Additionally, a comparison to the initial, relaxed, system is made. 
A time step of $0.4~\rmn{Myr}$ is used and the length of the integration corresponds to
1 Gyr.

\subsection{Results}

\begin{figure}
\centering
\includegraphics[width=0.47\textwidth]{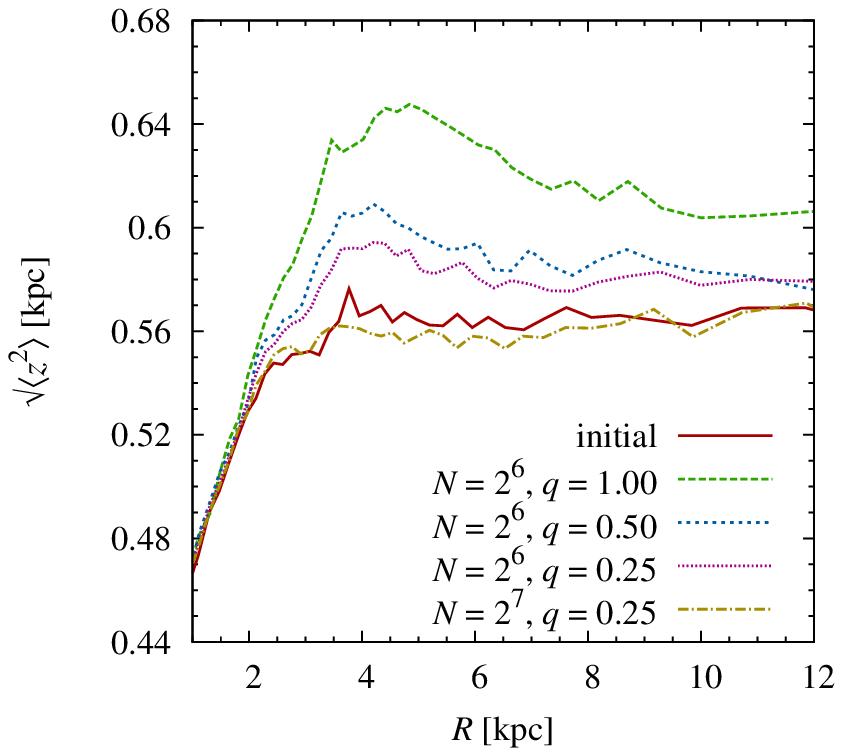}
\caption{Standard deviation of the $z$-coordinate of disc particles as function of $R$
for varying $N$ and $q$ at $t = 1~\rmn{Gyr}$, together with the initial curve. The values are
calculated in radial bins of equal particle numbers.}
\label{fig:z.stddev}
\end{figure}

Figure~\ref{fig:z.stddev} shows the root mean square (RMS) of the $z$-coordinate 
of all disc particles, $\sqrt{\langle z^2\rangle}$, as a function of radial distance $R$.
We used radial bins of equal particle numbers. The RMS can be
taken as a measure of the thickness of the disc.
The low-resolution simulation ($N=64$, $q=1$) over-estimates the 
thickness significantly. The maximum deviation from
the initial thickness is about 16 per cent.
Flattening the intermediate grid by a factor of 4, brings the 
thickness down to within about 4 per cent of the initial values, without increasing the 
computation time. The medium-resolution simulation ($N=128$, $q=0.25$) deviates at
most $\approx 1.6$ per cent from the initial values, but is closer than $1$ per cent for most
of the radial range.

\begin{figure}
\centering
\includegraphics[width=0.47\textwidth]{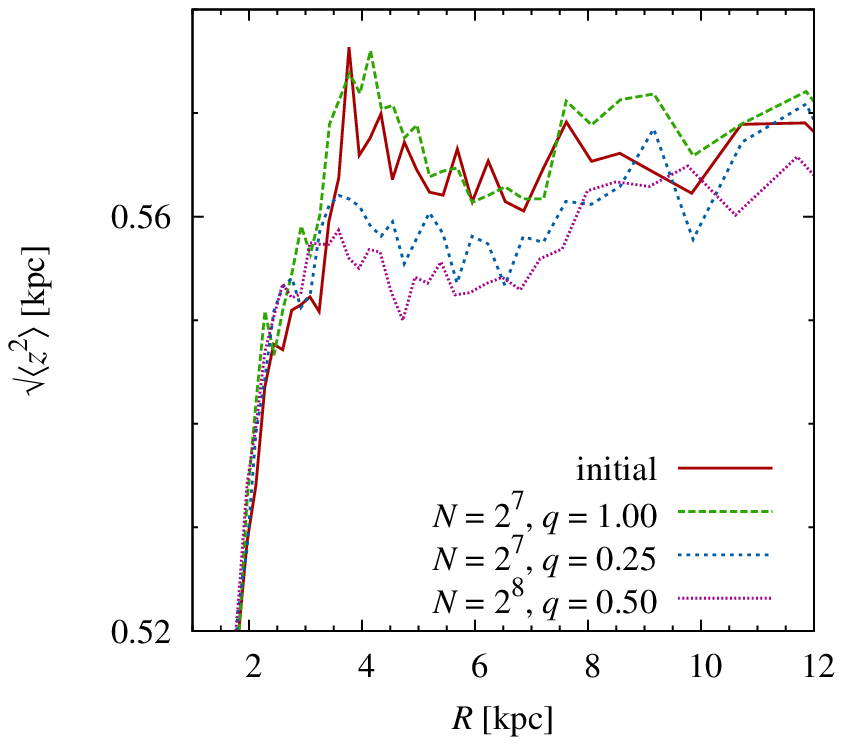}
\caption{Standard deviation of the $z$-coordinate of disc particles as function of $R$
for varying $N$ and $q$ at $t = 1~\rmn{Gyr}$, together with the initial curve. The values are
calculated in radial bins of equal particle numbers.}
\label{fig:z.stddev.comp}
\end{figure}

 To demonstrate the effect of flattening, Figure~\ref{fig:z.stddev.comp} directly
  compares two simulations with the same vertical resolution -- $N=128$ with $q = 0.25$
  and $N=256$ with $q=0.25$ -- corresponding to a vertical cell length of about 112 pc. 
  As a reference the case of $N=128$ without flattening is
also shown. As can be seen, introducing flattening in the case of medium resolution
diminishes thickness, down to about the same value as in the case of high resolution.

\begin{figure}
\centering
\includegraphics[width=0.47\textwidth]{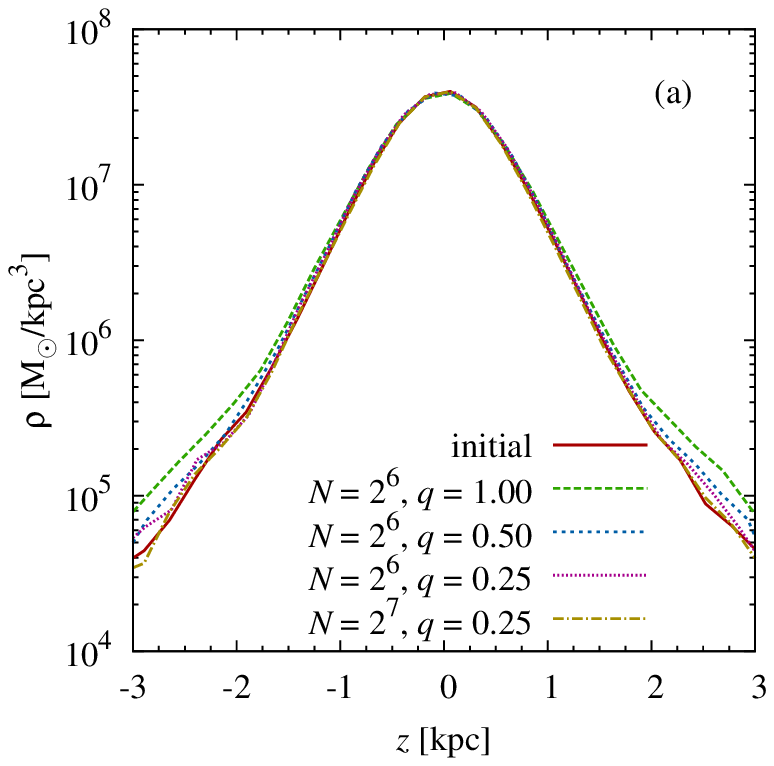}
\includegraphics[width=0.47\textwidth]{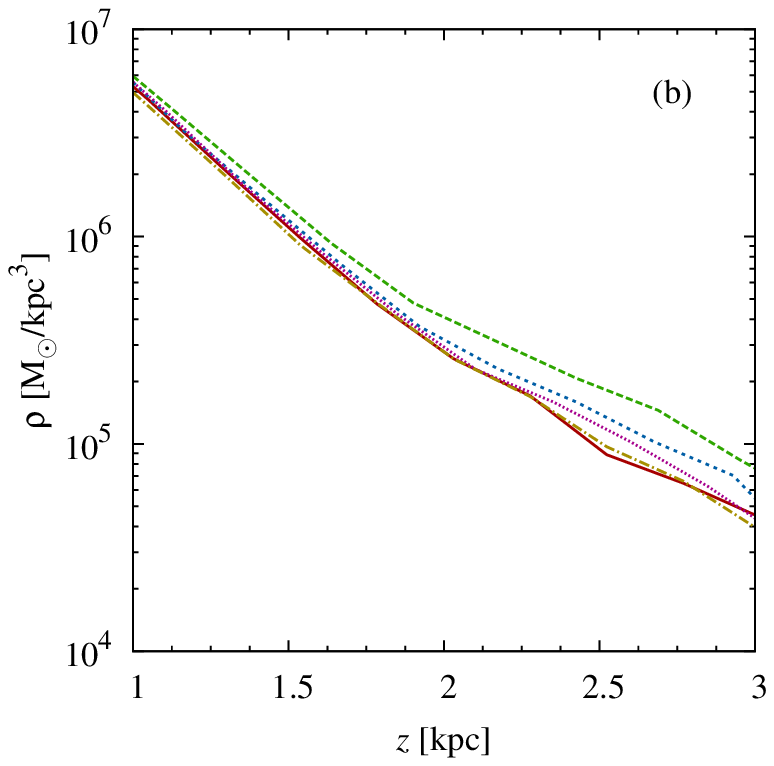}
\caption{Volume density $\rho$ as function of $z$ (averaged over the radial range 
$7.5~\rmn{kpc}\le R\le 8.5~\rmn{kpc}$) for varying $N$ and $q$ at $t = 1~\rmn{Gyr}$. 
The solid line shows the initial values. {\bf (a)} Whole range. {\bf (b)} Right tail.}
\label{fig:zprof}
\end{figure}

Figure~\ref{fig:zprof} displays the vertical density profile averaged
over the radial interval $7.5~\rmn{kpc} \le R \le 8.5~\rmn{kpc}$. 
The initial profile (solid line) and the profile of the medium-resolution simulation
(dash-dotted line) are almost identical. In the case of the
low-resolution simulations (dashed line), the profile is 
significantly widened for $|z|>1~\rmn{kpc}$ (see Figure~\ref{fig:zprof}~(b)). This is a
sign of numerical heating. Reducing $q$ brings the profile closer to the initial
one. For $q=0.25$, deviations are only visible beyond 2 kpc. 

\begin{figure}
\centering
\includegraphics[width=0.47\textwidth]{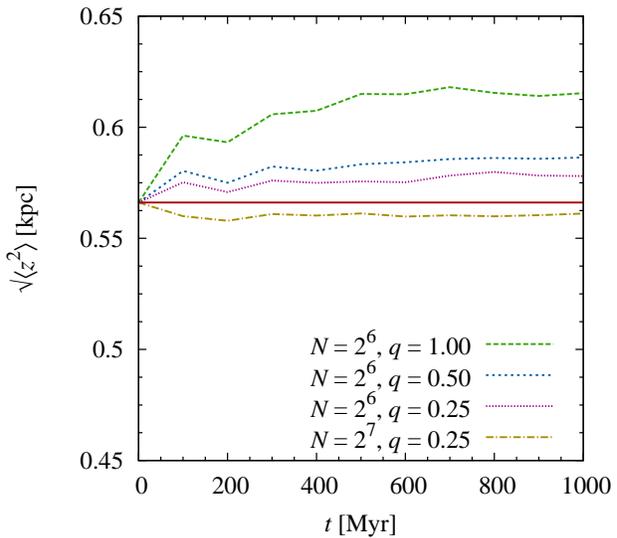}
\caption{Standard deviation of the $z$-coordinate of disc particles as function of
time for varying grid size $N$ and flattening $q$. 
The standard deviation is calculated in a region around $R = 8~\rmn{kpc}$. The solid line
shows the initial value.}
\label{fig:z.stddev.evo}
\end{figure}

Figure~\ref{fig:z.stddev.evo} shows the time evolution of the RMS
for varying grid size $N$ and flattening factor $q$ in the region around $R = 8~\rmn{kpc}$.
In the first 200 Myr, the model adapts to the new grid structure, which causes a
change in the RMS. After that, the RMS increases slightly with time in the case of the
low-resolution simulations. It reaches values that
are $8.7$ per cent ($q=1$), $3.6$ per cent ($q=0.5$) and $2.4$ per cent ($q=0.25$) greater 
than the initial
value. In the medium-resolution simulation however, it remains stable at a deviation of
about $0.9$ per cent.


\section{Merger simulations}
\label{sec:mergesim}

One application that benefits from the increased z-resolution
is the study of the dynamical heating of galactic discs 
caused by the merging with satellite galaxies. We simulate the merging
of a small satellite with a disc-bulge-halo galaxy for varying initial
positions and velocities of the satellite.

\subsection{Simulation parameters}

All satellites are initially at a distance of $R_{\rmn{A}}=25~\rmn{kpc}$ from the 
centre of the galaxy and have a velocity of either $115~\kms$ or $81.3~\kms$.
If all the mass of the galaxy inside $R_{\rmn{A}}$ were concentrated in a
point mass, these velocities would result in elliptical orbits with apocentre distance
$R_{\rmn{A}}$ and eccentricities
$\epsilon = 0.56$ and $\epsilon = 0.78$, respectively. Here, the eccentricity is defined
as
\[ \epsilon = {\sqrt{a^2-b^2} \over a}\]
with semi-major axis $a$ and semi-minor axis $b$. In the two-body problem,
the pericentre distance of these orbits would be $R_{\rmn{P}} \approx 7~\rmn{kpc}$ and
$R_{\rmn{P}} \approx 3~\rmn{kpc}$, respectively.

\begin{table}
\centering
\caption{Satellite parameters $\epsilon$ (eccentricity), $v$ (velocity), $R_{\rmn{A}}$ (apocentre),
$R_{\rmn{P}}$ (pericentre) and $i$ (orbital inclination).}
\label{tab:satparam}
\begin{tabular}{@{}c|cccc}
\hline
$\epsilon$ & $v~[\rmn{km}/\rmn{s}]$ & $R_{\rmn{A}}~[\rmn{kpc}]$ & $R_{\rmn{P}}~[\rmn{kpc}]$ & $i~[\rmn{\degr}]$ \\\hline
$0.56$ & $115$ & 25 & 7.05 & 0, 10, $\dots$, 180\\
$0.78$ & $81.3$ & 25 & 3.08 & 0, 10, $\dots$, 120\\\hline
\end{tabular}
\end{table}

In addition to the eccentricity, we vary the orbital inclination $i$ relative to the plane
of the disc between $0\degr$ and
$180\degr$ for $\epsilon = 0.56$ in steps of $10\degr$, and likewise between $0\degr$ and $120\degr$ for $\epsilon=0.78$.
For $i < 90\degr$ the satellite is on a prograde orbit as compared to the galactic rotation,
while for $i > 90\degr$ the orbit is retrograde. For $i=90\degr$ the satellite's initial velocity
is perpendicular to the bulk motion of the disc.
Table~\ref{tab:satparam} summarises these parameters.

As shown in the previous section, a grid cell number of $N=128$ with flattening
factor $q=0.25$ causes no significant numerical heating. We adopt these
values for our merger simulations.
All simulations run for 1 Gyr. After that, the satellites are completely dissolved.
As a control, the galaxy model is also evolved in isolation.

At the end of the simulation, the plane of the disc is tilted by a few degrees. 
In the following, the data are evaluated in a coordinate system where $x$
and $y$ define the galactic plane and $z$ is perpendicular to it.

\subsection{Vertical profile}

\begin{figure}
\centering
\includegraphics[width=0.47\textwidth]{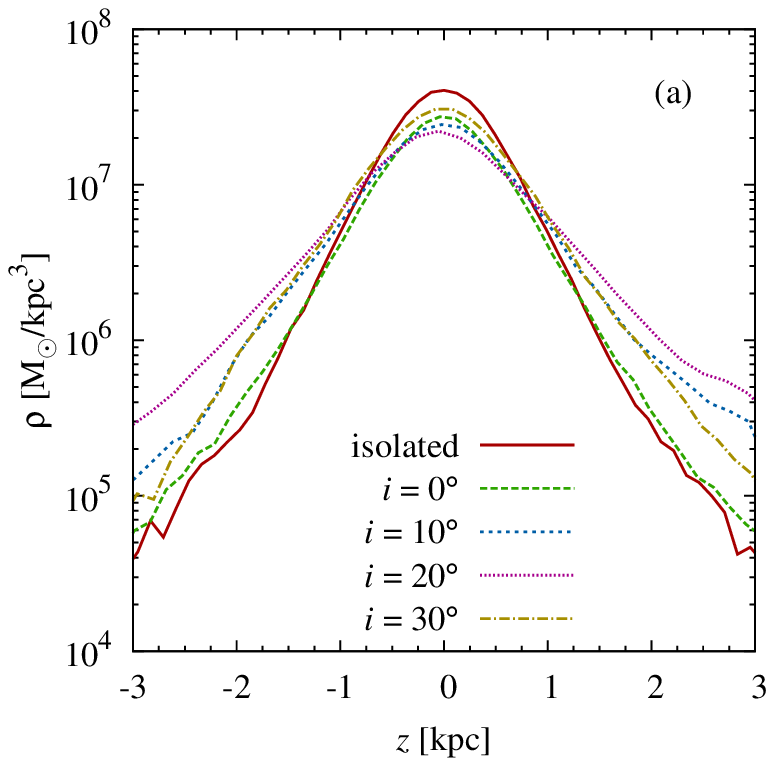}
\includegraphics[width=0.47\textwidth]{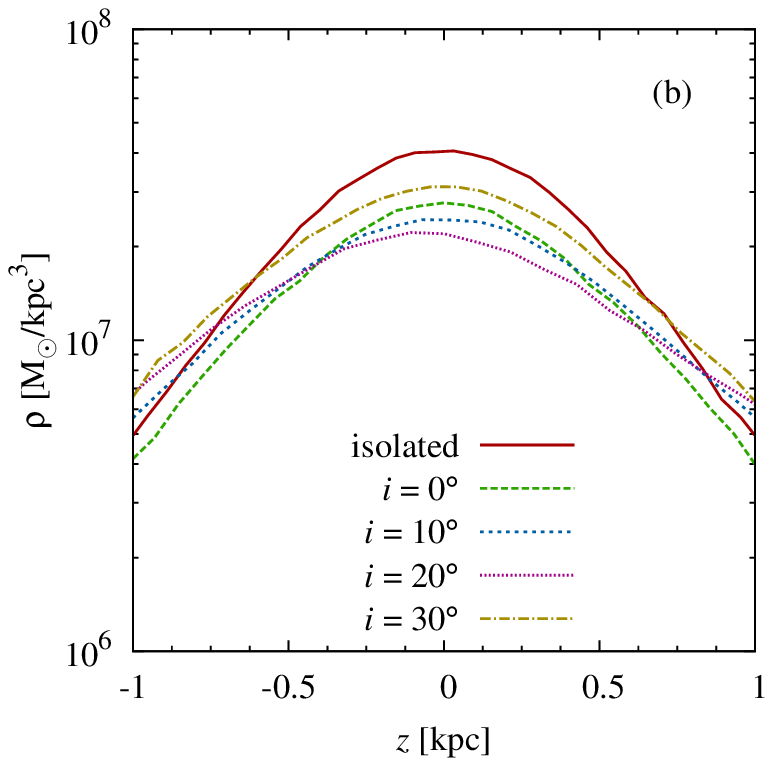}
\caption{Volume density of the disc in the region $7.5~\rmn{kpc}\le R\le 8.5~\rmn{kpc}$
after the merging ($t=1~\rmn{Gyr}$) as function of $z$ for different satellite orbital inclinations $i$.
All satellites show have initial orbital eccentricity $\epsilon = 0.56$. The solid line corresponds
to the isolated simulation, i.e., without satellite.
{\bf (a)} Whole range. {\bf (b)} Inner part.}
\label{fig:merge-zprof}
\end{figure}

Figure~\ref{fig:merge-zprof} shows the vertical density profile of the disc
after merging. Only satellites with significant effect on the disc
are shown ($e = 0.56$, $i = 0\degr\dots 30\degr$).
Compared to the isolated simulation (i.e. without satellite), the central density is decreased 
by a factor of 2 at the most ($i = 20\degr$).
In the inner part ($|z|<1~\rmn{kpc})$, the original $\rmn{sech}^2(z/z_0)$ profile
remains a good fit, albeit with a greater thickness $z_0$. The isolated galaxy has
$z_0 = 0.6~\rmn{kpc}$, while for $i = 20\degr$ the thickness is increased to 
$0.8~\rmn{kpc}$. 
In the outer part, however, the density is increased as compared to a $\rmn{sech}^2(z/z_0)$ law.
While the thickness of $0.8~\rmn{kpc}$ corresponds to a scale height of $0.4~\rmn{kpc}$,
the outer part is better fitted by an exponential function with scale height $\approx 0.65~\rmn{kpc}$.

  \cite{hc06} provide a formula for the increase of the thickness of the disc caused by cold
  dark matter subhaloes, which should also be applicable to satellite mergers. 
  In our notation their formula is given by
  \[ \Delta z_0 \approx 8 h \left(\frac{M_{\rmn{sat}}}{M_{\rmn{disc}}}\right)^2 \approx 0.25~\rmn{kpc}~\rmn{.}\]
  This is compatible with an increase of $\Delta z_0\ga 0.2~\rmn{kpc}$ in the case of $i = 20\degr$ 
  as described above.

\subsection{Heat increase}

We define the total heat in the disc as the kinetic energy of random motion
\begin{equation}
  E_{\rmn{heat}} = \sum_i {1 \over 2} m_i \left|\mathbf{v}_i-\overline{\mathbf{v}}_c\left(R_i\right)\right|^2
\end{equation}
where the sum goes over all particles in the disc, $m_i$ is the mass of the $i$th particle,
$\mathbf{v}_i$ is the velocity of the $i$th particle,
and $\overline{\mathbf{v}}_c\left(R_i\right)$ is the mean circular velocity at the radial
distance $R_i$ of the $i$th particle. $\bmath{v}_c$ is first calculated in radial
bins of equal particle numbers and then interpolated to the individual distances $R_i$.

\begin{figure}
\centering
\includegraphics[width=0.47\textwidth]{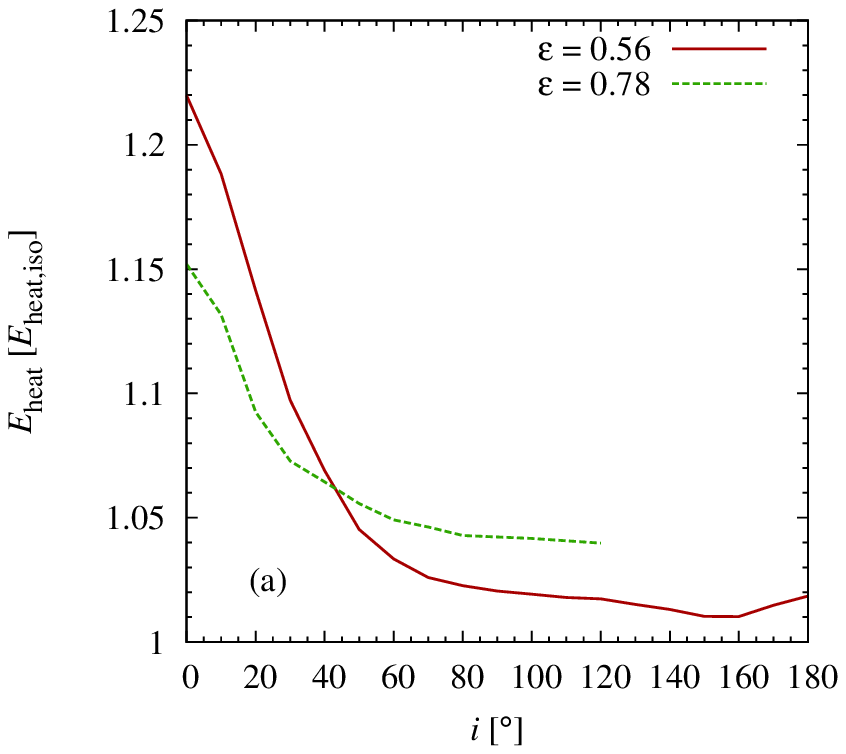}
\includegraphics[width=0.47\textwidth]{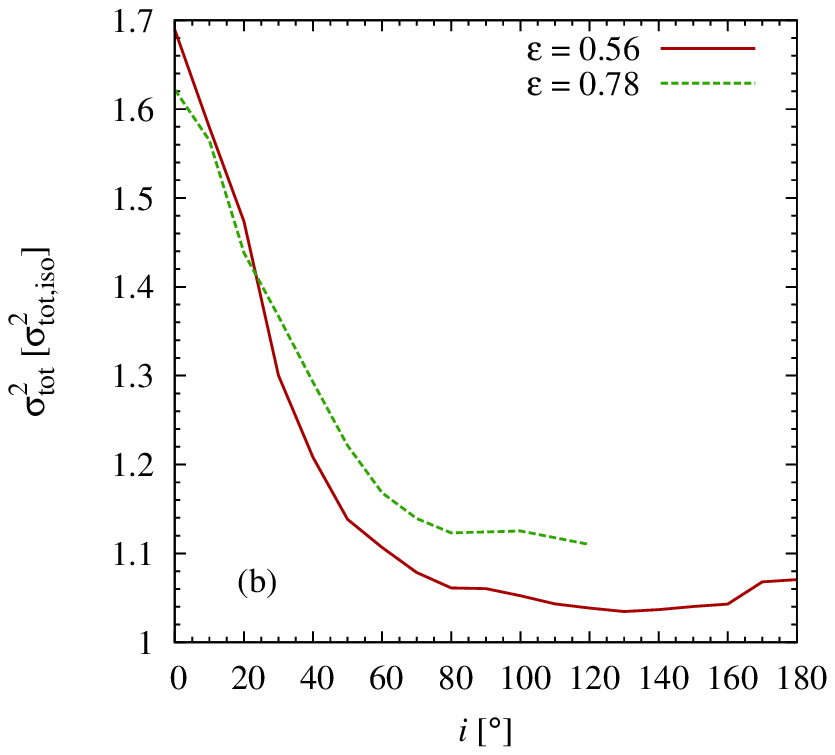}
\caption{{\bf (a)} Total heat of the disc after merging ($t=1~\rmn{Gyr}$) as a
function of satellite's orbital inclination $i$ for varying eccentricity $\epsilon$.
$E_{\rmn{heat}}$ is measured in units of the heat of the isolated galaxy, 
$E_{\rmn{heat, iso}}$, at $t=1~\rmn{Gyr}$.
{\bf (b)} Squared total velocity dispersion at $t = 1~\rmn{Gyr}$ in the region around 8 kpc as a
function of orbital inclination $i$ and for varying eccentricity $\epsilon$.
$\sigma_{\rmn{tot}}^2$ is measured in units of the squared total velocity dispersion of the
isolated galaxy.}
\label{fig:heat}
\end{figure}

Figure~\ref{fig:heat} (a) shows $E_{\rmn{heat}}$ 
in units of the heat in the isolated disc $E_{\rmn{heat,iso}}$ at the end of the simulation 
(i.e. at $t = 1~\rmn{Gyr}$).
$E_{\rmn{heat,iso}}$ increases only by $0.18$ per cent as compared to the initial value,
demonstrating
that our simulations are free of numerical heating. Figure~\ref{fig:heat} (b) displays
the same for the square of the total velocity dispersion in the solar neighbourhood.

The heating efficiency of
the satellite depends on its orbital inclination as well as on its orbital eccentricity.
For low inclinations, satellites on less elliptical orbits heat more effectively than 
those on more elliptical ones. For $i > 40\degr$, however, satellites on orbits
with $\epsilon = 0.78$ are more effective.
Prograde satellites heat the most with an maximum increase of 22 per cent, while 
retrograde ones only have a small effect of about 2 per cent. These results 
are consistent with those obtained by \cite{vw99}.

Observations show that the velocity dispersion in the solar neighbourhood
increases from about 30 $\kms$ to about 60 $\kms$ in 8 Gyr  
approximately proportional to $\sqrt{t}$ \citep[see, e.g.,][]{hf02}. 
This means, that the specific heat
increases linearly by about $1350~\rmn{km}^2~\rmn{s}^{-2}$. If we assume that 
the heat increase is completely due to satellite mergers, then the required
merger rate $n$ is
\begin{equation}
  \nu = \frac{1}{\Delta h}\times \frac{1350}{8}\frac{\rmn{km}^2}{\rmn{s}^2~\rmn{Gyr}}
\end{equation}
where $\Delta h$ is the specific heat increase imparted by a single merger.

In our simulations,
we found that for low inclinations the heat increase is approximately uniform
over the whole disc (except for the bulge-dominated inner part) and lies
between 100 and 1000 $\rmn{km}^2~\rmn{s}^{-2}$ per merger. For high inclinations, the
heat increase is mainly due to flaring of the outer parts of the disc ($R > 15~\rmn{kpc}$),
while the solar neighbourhood is basically unaffected. Only considering low
inclinations, the required merger rate then lies between $\nu=0.17$ and $\nu=1.69$ mergers per Gyr, 
depending on inclination and eccentricity of the orbit.

\subsection{Velocity dispersion ratio}

\begin{figure}
\centering
\includegraphics[width=0.47\textwidth]{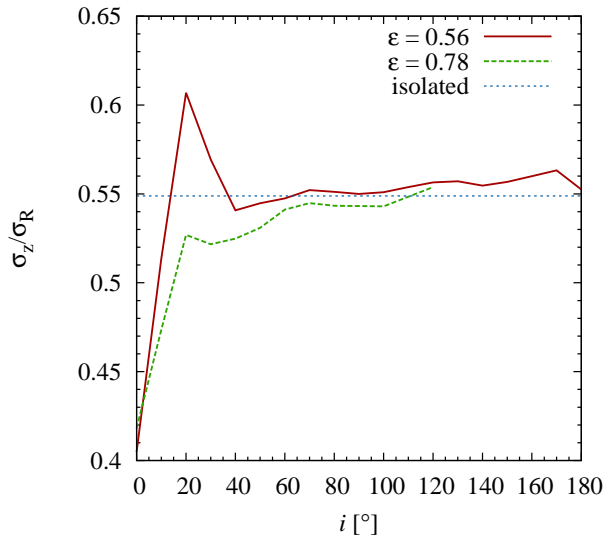}
\caption{Velocity dispersion ratio $\sigma_z/\sigma_R$ at $t = 1~\rmn{Gyr}$ in the 
region around 8 kpc as function of orbital inclination $i$ and for varying eccentricity $\epsilon$.}
\label{fig:sigrat}
\end{figure}

Observations show that the current ratio of the radial and vertical velocity dispersions
in the solar neighbourhood, $\sigma_z/\sigma_R$, is approximately $0.5\pm 0.1$ and 
increases slightly with stellar age, $\propto t^{0.16}$ \citep{h07}.
Figure~\ref{fig:sigrat} shows this ratio after
merging as a function of orbital inclination $i$.
The satellites with low prograde orbits decrease the ratio to at most about 0.4, i.e., 
they heat more efficiently in radial than in vertical direction. In one case ($i=20\degr$,
$\epsilon=0.56$) there is an increase to approximately 0.6.
For inclinations above $30\degr$ the ratio remains mostly unchanged in the solar neighbourhood.

\subsection{Energy and angular momentum transfer}

An interesting question is where the initial energy and angular momentum of the
satellite end up in the final system.

\begin{figure}
\centering
\includegraphics[width=0.46\textwidth]{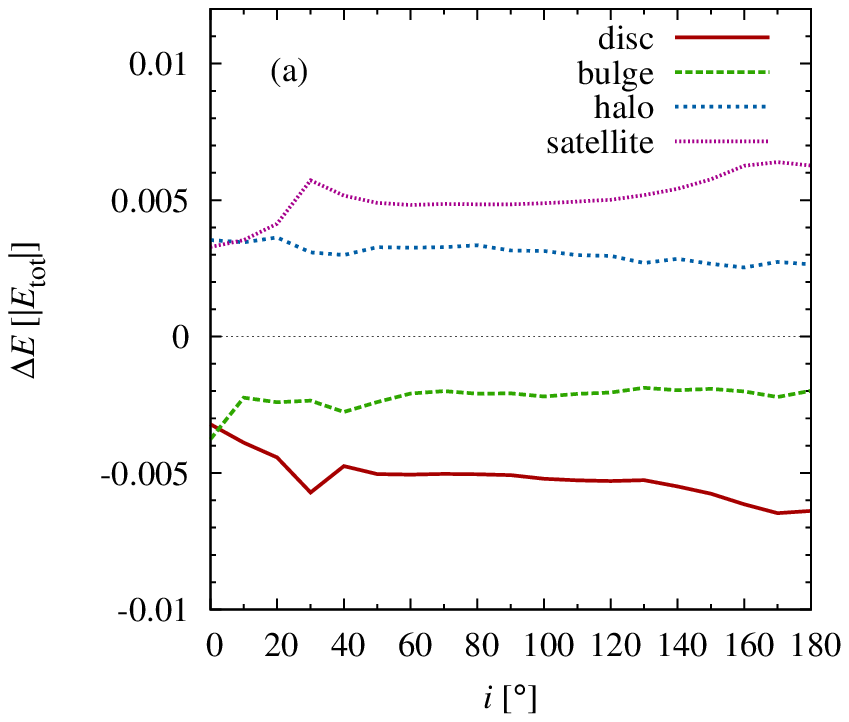}
\includegraphics[width=0.46\textwidth]{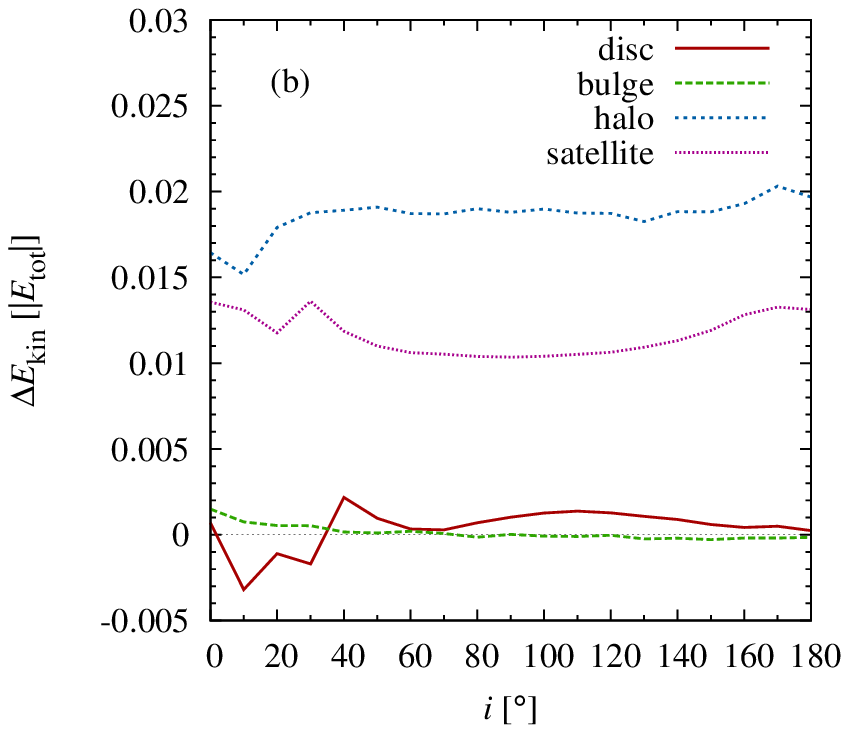}
\includegraphics[width=0.46\textwidth]{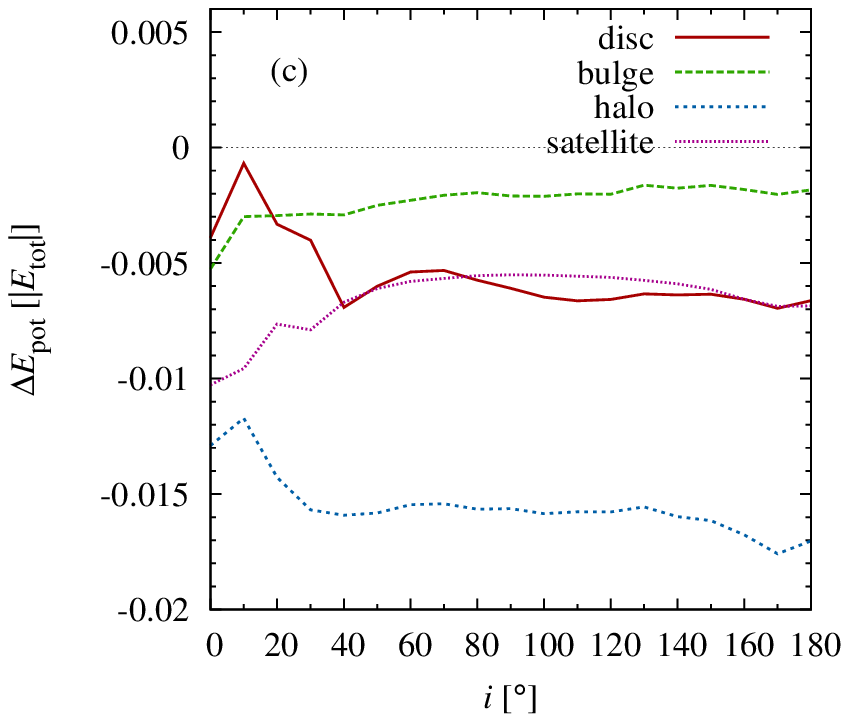}
\caption{Change in total (a), kinetic (b) and potential (c) energy of the four
components with respect to the initial energies as a function of orbital inclination $i$. 
The energy change is measured in units of the total energy of the isolated galaxy. 
Only the data for $\epsilon = 0.56$ is shown.}
\label{fig:energy-transf}
\end{figure}

Figure~\ref{fig:energy-transf} shows the change in energy of the various components.
The energy is
measured in units where the total energy of the isolated galaxy is $-1$. ``Satellite'' designates
those particles that once made up the satellite, but now
are distributed over the whole system.

The initial galaxy-satellite system is not in equilibrium.
As a consequence of virialisation, both the satellite and the halo gain 
kinetic and lose potential energy. This results in a deeper potential well
in the centre, which reduces the potential energy of bulge and disc.
During its in-spiral, the satellite transfers part of its kinetic energy 
to the halo particles. This is due to dynamical friction.

\begin{figure}
\centering
\includegraphics[width=0.47\textwidth]{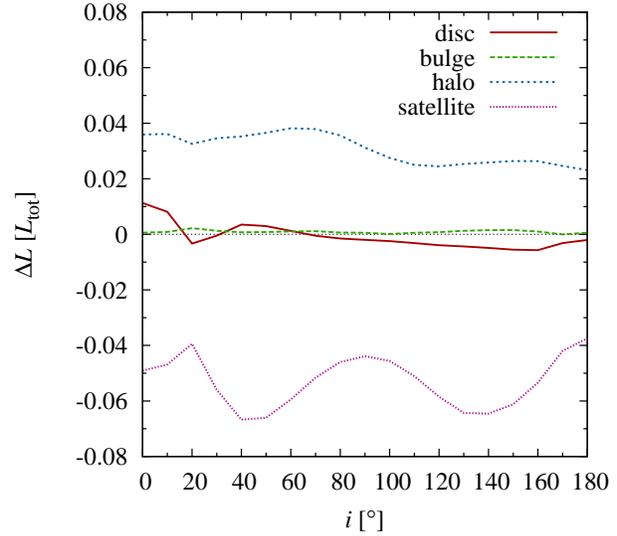}
\caption{Change in total angular momentum of the four components with respect to the initial
angular momenta as a function of orbital inclination $i$. The change is measured in units 
of the total angular momentum of the isolated galaxy. Only the data for $\epsilon = 0.56$ is shown.}
\label{fig:angmom-transf}
\end{figure}

This effect can also be seen in Figure~\ref{fig:angmom-transf},
which shows the change in angular momentum as a function of inclination, measured
in units of the total angular momentum of the isolated galaxy. The angular
momentum of the bulge barely changes. This is not surprising since the satellites are already
mostly destroyed before reaching the inner part of the galaxy. 
Our data reveal that the radial distribution of angular momentum in the disc changes: 
the disc slows down and expands. Its {\em total} 
angular momentum, however, remains approximately nearly constant. It only increases slightly for
satellites on prograde orbits and decreases slightly for satellites on retrograde orbits.
The satellite's initial orbital angular momentum is $0.31$ in
these units. Accordingly, it loses between 15 per cent and 20 per cent of its angular momentum.
Of that, about 80 per cent goes into the halo, while the rest is imparted onto the disc.

\subsection{Comparison with higher resolution}

To ensure that the choice of resolution does not influence the results we
simulate a case with $\epsilon = 0.56$ and $i = 0\degr$, and higher resolution.
We choose $N=256$ and $q=0.125$, as this is the highest possible resolution 
one can achieve on our present hardware.

We calculate the Fourier transform of the density distribution of the disc 
at the end of the simulation in order
to find possible perturbations caused by low resolution. The $m$th complex 
Fourier coefficient is given by

\[
  \tilde{A}_m(R_i) = \frac{1}{S_i} \sum_j M_j e^{-im\varphi_j}
\]

where $R_i$ and $S_i$ are the central radius and area of the $i$th bin, the index
$j$ runs over all particles in bin $i$, and $M_j$ and $\varphi_j$ are the mass and
polar angle of the $j$th particle. The amplitude of $\tilde{A}_m$ is denoted by
$A_m$ and the phase by $\theta_m$.

\begin{figure}
\centering
\includegraphics[width=0.47\textwidth]{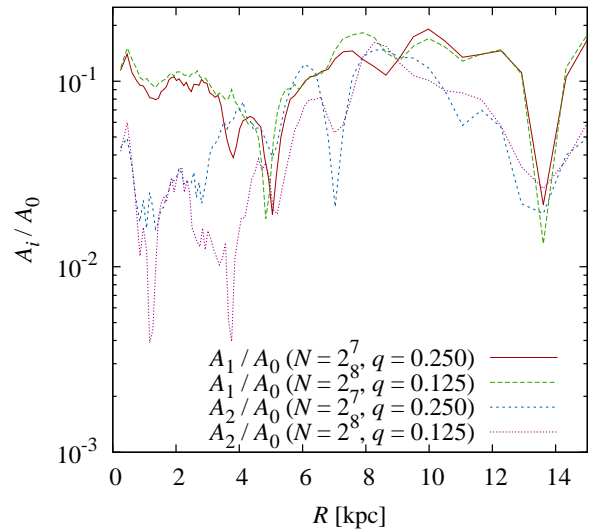}
\caption{Amplitudes of the first and second spiral mode as a function of radius
  at the end of the simulation.}
\label{fig:modes}
\end{figure}

\begin{figure}
\centering
\includegraphics[width=0.47\textwidth]{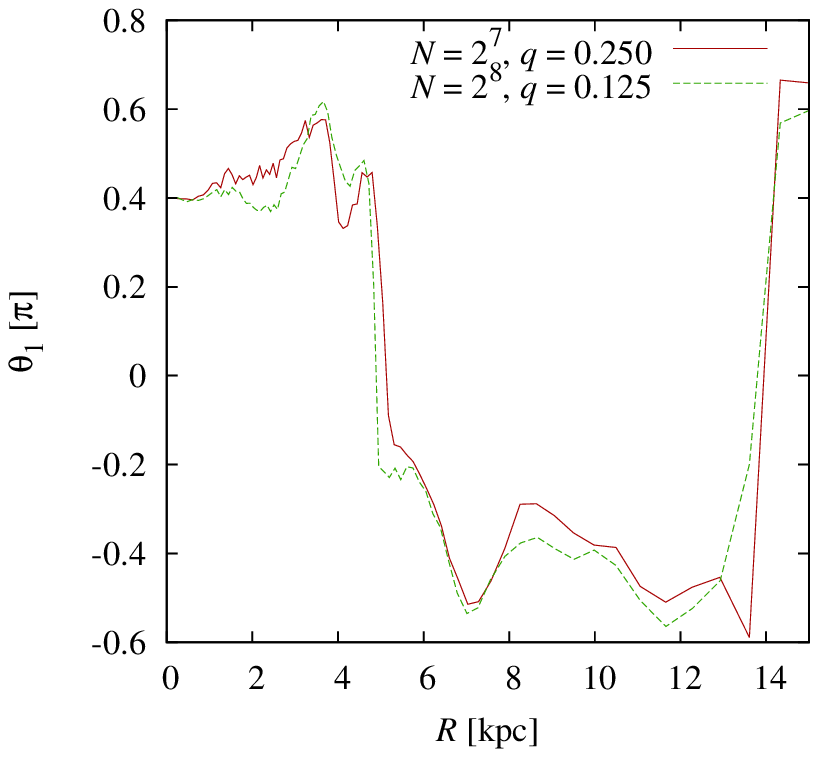}
\caption{Phase of the first mode as a function of radius
  at the end of the simulation.}
\label{fig:phases}
\end{figure}

Figure~\ref{fig:modes}
shows the amplitudes of the first and second mode normalised by the mean density,
while Figure~\ref{fig:phases} displays the phase of the first mode.
The two resolutions result in a roughly similar distribution in Fourier space. 
There are no strong perturbations in amplitude caused by the lower resolution. 
Neither is the pattern speed affected significantly.


\section[]{Conclusions}

{\sc Superbox} is a particle-mesh code where additional grids and sub-grids are 
applied to regions of high particle density. This strategy turns out to be very efficient
since the code can run on any workstation or PC. Nevertheless, the code has its
limitations. For instance, stellar discs are poorly resolved in vertical 
direction. We overcome this problem by introducing flattened grids. This is one
of the features of the new code {\sc Superbox-10} where, in addition, an individual
mass can be assigned to each particles.

We found that the computationally most intensive part of the code is the FFT.
We parallelised it using the library {\sc fftw}. This resulted in a
speed-up of 2 to 24, depending on grid size $N$ and the number of particles $n$.

We created a galaxy model with an exponential disc, a bulge with a Hernquist
profile and a dark matter halo, also with a Hernquist profile. The model was
realised using the program {\sc MaGaLie}. 

The model was first simulated in isolation. These simulations show that flattening 
the intermediate grid is an efficient means to reduce numerical heating in the simulation.

We also simulated the merging of the galaxy with small satellites in order to analyse the 
proposed disc heating due to the interaction.
We find that satellites on prograde orbits with low eccentricity
and inclination heat the disc most efficiently. If the heat increase in
the solar neighbourhood were to be explained by that type of satellite mergers alone, a 
rate between 0.2 and 1.7 mergers per Gyr would be
required. The detailed analysis of energy and angular momentum re-distribution shows that 
most of the satellites energy and angular momentum is transfered to the dark matter halo.
This shows that the halo plays an important role even in 50:1 mergers. This confirms, 
that simulations of such processes should represent the halo by live particles and not by 
a fixed background potential. The presented pilot study of high-resolution simulations of 
disc heating by merging satellite galaxies serves as a starting point for an extended 
parameter study to quantify the heating rate in a cosmological context.

\section*{Acknowledgments}

We thank Markus Hartmann und Ingo Berentzen for
numerous and fruitful discussions.

This work was supported by Sonderforschungsbereich SFB 881 ``The Milky Way System'' 
(subproject A1) of the German Research Foundation (DFG).

The computer hardware used for the simulations was supported by project ``GRACE''
I/80 041-043 of the Volkswagen Foundation and by the Ministry of Science, Research
and the Arts of Baden-W\"urttemberg.


\bsp

\label{lastpage}

\end{document}